\documentclass[aps, amsfonts, twocolumn, nofootinbib,notitlepage, preprintnumbers,raggedbottom]{revtex4}
\usepackage{hyperref}
\hypersetup{
colorlinks=true,
linkcolor=red,
citecolor=blue,
urlcolor=blue}
\usepackage{epsf,epsfig}
\usepackage{amscd}
\usepackage{bbold}
\usepackage{amsmath}

\input xy
\xyoption{all}

\renewcommand{\eqref}[1]{Eq.~(\ref{#1})}

\newcommand{\Smatrix}{$S$-matrix}
\newcommand{\Smatrices}{$S$-matrices}
\newcommand{\T}{{\sffamily T}}
\newcommand{\lbh}{\ensuremath{\frac{L}{2}}}
\newcommand{\oneover}[1]{\ensuremath{\frac{1}{#1}}}
\newcommand{\half}{\oneover{2}}

\newcommand{\goesto}{\ensuremath{\rightarrow}}

\newcommand{\Luscher}{L\"uscher}
\newcommand{\asy}{\ensuremath{ _{\text{asy}}}}
\newcommand{\BZL}{\ensuremath{B^Z_L}}

\newcommand{\va}{\ensuremath{\vec{\alpha}}}

\newcommand{\nn}{\nonumber}

\newcommand{\ket}[1]{\ensuremath{\left|\;#1\;\right\rangle}}										
\newcommand{\bracket}[2]{\ensuremath{\left\langle\;#1\;\middle|\;#2\;\right\rangle}}				
\newcommand{\braket}{\bracket}																		

\begin{document}

\title{Multi-channel \Smatrices\ from energy levels in finite boxes}
\author{Evan Berkowitz}\email{evanb@umd.edu}
\author{Thomas D. Cohen}\email{cohen@umd.edu}
\author{Patrick Jefferson}\email{patrick.c.jefferson@gmail.com}
\affiliation{Maryland Center for Fundamental Physics\\ 
Department of Physics\\
University of Maryland, College Park, MD USA\\
\today}

\preprint{UM-DOE/ER/40762-526}

\begin{abstract}
We show that for a generic quantum mechanical system with more than one open scattering channel, it is not possible to fully reconstruct the theory's \Smatrix\ from spectral information obtained in large finite volumes with periodic boundary conditions.  Physically distinct  \Smatrices\ can have identical finite-volume spectra for large finite boxes of arbitrary sizes.  If the theory is not time-reversal symmetric, there exists an uncountably infinite set of distinct \Smatrices\ with the same spectra.   If the theory respects time-reversal symmetry there exists a discrete set of  \Smatrices\ with identical energy levels for finite boxes.  We illustrate the issue for simple quantum mechanical systems in 1+1 dimensions.  
\end{abstract}
\maketitle

\section{Introduction}

Because it is strongly coupled at low energy, quantum chromodynamics (QCD) is not amenable to perturbative techniques there.  Fortunately, lattice simulations provide a first-principles technique for extracting predictions from the theory.  The lattice has achieved unprecedented accuracy for many QCD observables, including some that are difficult to extract experimentally\cite{lattice2011}.  Indeed, it has become a precision tool for many purposes.  However, not all QCD observables are easily accessible via lattice simulations.  One important class of these are scattering observables in regimes where there is more than one open channel.  

Before early groundbreaking work\cite{Hamber198399,luscher:1986I,luscher:1986II,wiese1989,Luscher1991,Luscher1991237} it was not generally understood that at least some types of  scattering information are accessible via finite-volume lattice studies, which by construction always yield a discrete spectrum of energy levels.  Now it is well-understood that in cases where the kinematics restrict the scattering to a single two-body scattering channel, one can extract the phase-shift for the channel given knowledge of the energy levels in a finite box.  In particular given a sufficiently large box, knowledge of the size of the box, and the eigenenergies in that box, one can extract phase-shifts for those energies.  By scanning through different box sizes---and thus different energies---one can extract the phase shift as a function of energy.  This approach has already proven to be practical for cases of low-energy scattering\cite{alexandru:2012,PhysRevD.83.094505,prelovsek:2011,PhysRevD.84.094505}. However, this  technique, usually referred to as ``\Luscher's formula'', only provides access to certain kinds of data.  It   does not apply in its original form to situations in which there is more than one accessible channel.  

This difficulty is acute for the problem of hadron spectroscopy, a subject in which there has been considerable interest from the lattice community  \cite{Guadagnoli200474,Dudek:2007wv,Basak:2007kj,Bulava:2009jb,Dudek:2009kk,Dudek:2009qf,Mahbub2009418,Dudek:2010wm,Bernard:2010fp,Edwards:2011jj,Liu:2012ze,PhysRevD.83.094505}.   The difficulty is that the vast majority of hadrons are resonances above the three-body threshold for which the original formulation of the \Luscher\ formula is inapplicable.  The current state of the art for high-lying resonances is simply to calculate in a given relatively small box size and to identify the resonance's mass with an energy level.  Such a procedure is presumably a valid way to  identify the position of the resonance in the limit of an infinitely narrow resonance and in a regime where the box size is large enough to avoid large finite volume effects.  However, the hadrons of physical interest typically  have fairly substantial widths so this procedure is questionable.  Moreover, the ``mass'' of a resonance is a quantity which depends on how one separates peaks from background in scattering data; such separation inherently creates model dependency.  It would be far preferable to have access directly to the scattering observables in the form of the \Smatrix\ directly from the lattice.    This would enable one to make contact with the physics associated with hadronic widths and branching ratios.   

The need for going beyond a simple association of a hadron's mass with the energy of a resonance is particularly clear in cases where the width of a resonance is large and multiple channels are involved.  A good example is the Roper resonance: although it has a low excitation energy, it is quite broad and is strongly coupled to the nucleon plus two pion channel.  The Roper resonance is surprisingly difficult to characterize experimentally and the width and branching ratios are poorly determined\cite{PDG}.  Thus for the Roper resonance, where an {\it ab initio} calculation of the scattering observables would be particularly helpful in trying to understand the nature of the  state, a direct extraction of the mass from the energy on a  single finite volume lattice is particularly difficult. 

There are other cases where a direct extraction of the \Smatrix\ is important.  For example, in the on-going program to attempt to understand nuclear phenomena from QCD, low energy nuclear scattering observables are of interest.  The $^1S_0$ channel in nucleon-nucleon scattering involves a single channel; it's \Smatrix\ is fully characterized by a phase shift and the  \Luscher\ formula directly applies.  However,  scattering in the deuteron channel is coupled and includes mixing between the $^3S_1$ and  $^3D_1$ partial waves and the  \Luscher\ formula  does not directly apply.  To proceed with the program of extracting nuclear physics from lattice  QCD  it is important to understand how to extract the \Smatrix\ from the lattice.  

There have been significant  recent theoretical developments aimed at generalizing the \Luscher\ formula relating the energy levels in a finite box with periodic boundary conditions to many-channel  \Smatrices \cite{Polejaeva:2012,Lage:2009zv,Hansen:2012tf,Bernard:2012}.  Underlying these works is the key insight behind the \Luscher\ formula: when the box is significantly larger than the range of interactions between the particles, then the particles are essentially non-interacting when they encounter the boundary conditions associated with the box.  Thus the \Smatrix\  alone determines whether the boundary conditions are met.  One expects  that complete knowledge of the \Smatrix\ will determine the energy levels in the box.

This paper concentrates on the inverse problem:  If one has access to the energy levels of a system computed on a set of finite boxes with periodic boundary conditions,  can one reconstruct the full multi-channel \Smatrix?  This is the nature of the problem presented by lattice QCD.  In the case of \Luscher's formula, in which only a single channel is relevant,  this is straightforward.  The existence of a bound state at a given energy is fixed by a single piece of information---the phase shift at that energy---which can be read off directly given knowledge of the box size.   By working at a variety of box sizes which act to change the energies, one  can then map out the phase shift as a function of energy.  This is tantamount to learning the entire \Smatrix.   However in the multi-channel case, the \Smatrix\ contains more than a single piece of information at one energy.  It is thus not possible to determine a multi-channel \Smatrix\  at a given energy from the knowledge of levels in a single finite box---even if the discrete energy levels in the box include the energy of interest.  

Given this situation, it is immediately apparent that if it is possible to reconstruct the full \Smatrix\ at all, it will necessarily involve extracting the energy levels from more than a single box.  However, it is not immediately obvious whether it is possible to extract the  \Smatrix\ at a given energy even from the knowledge of energy levels in many different boxes.  Note that, in the case of a single channel problem where the \Luscher\ formula holds, if one has two different box sizes which share a common energy level, they give redundant information: both boxes allow for the extraction of the same phase shift at that energy (up to  finite volume corrections).  It is by no means clear  that in a multi-channel situation, where many parameters are needed to specify the \Smatrix, that different box sizes will also not give some redundant information in such a way as to prevent one from fully recovering the  \Smatrix.   Indeed we will show that in general it is not possible, even in principle, to fully reconstruct the  \Smatrix\ at some given energy, even with a knowledge of the spectrum for an arbitrarily large number of distinct box sizes.  In the general case, there exist physically distinct \Smatrices\  which yield the same same spectra in any finite box subject to periodic boundary conditions.  Fortunately, it may turn out that the physical observables of immediate relevance to two-body scattering can be reconstructed, albeit with a significant amount of work, and will require the study of sources with non-zero total momentum.

This paper is aimed at clarifying issues of principle.  Given this, it is useful to study the simplest context for which the issues arise.  To do this we will make several important departures from the case of greatest interest---a system with 3+1 space-time dimensions, with fully relativistic kinematics and with any number of particles in the final state. 

Instead, we focus on systems with 1+1 space-time dimensions.  The full 3+1 dimensional problem is much more complicated technically.  This complication arises from the fact that the free space theory is rotationally invariant, giving rise to a partial wave decomposition for the \Smatrix, while the finite box only has a discrete subgroup of the rotations associated with the symmetries of a cube.  However,  it is plausible that the answers to the most fundamental questions of principle are not likely to depend on the number of space-time dimensions.  It seems almost  certain that if it turns out not to be possible to extract \Smatrices\ from the finite volume data for 1+1 dimensional systems then it will not be possible for 3+1 dimensional systems. Thus, as a  first step it is  sensible to illustrate the impossibility of specifying the \Smatrix\ using a simple 1+1 dimensional system. 

In considering multi-channel \Smatrices\ there are two qualitatively different kinematic regimes to consider.  In one case, all of the channels involve two particles in the final state.  In the case of QCD one might consider scattering between a pion and a strange baryon ($\Lambda$ or $\Sigma$) in the $I=1$ channel in the regime where both the $\pi$-$\Lambda$ and $\pi$-$\Sigma$ channels are kinematically accessible but below the threshold for the $\pi$-$\pi$-$\Lambda$ channel.  Thus, while there are multiple channels in such cases the \Smatrix\ at any fixed energy can be fully specified by a discrete set of parameters.  The other case is when the system is in a kinematical regime where there may be more than two particles in the final state.  In this second case one needs to specify continuous functions to fully describe the \Smatrix\ at a fixed energy.  This second case is clearly more challenging.   Again, our aim here is to keep things as simple as possible to illuminate the underlying issues.   Thus, we will focus in this work on the first case---multiple channel problems with only two-particle final states.

Finally, the analysis of any particular problem depends on kinematical details.  These simplify for the case of nonrelativistic kinematics.  In the models studied in detail here we will restrict our attention to nonrelativistic systems, but our result easily generalizes to relativistic circumstances.

\section{A generalized \Luscher\ formula for a simple toy problem}
\label{sec:simple}

Consider the following nonrelativistic two-channel scattering problem in 1+1 space-time dimensions.  Suppose that there are two possible channels; each one consists of two identical bosons of mass $M$ but the particles in each of the two channels are different.  These bosons should be considered to be composite objects with finite size---as mesons are in QCD.  The energy in channel A is simply the kinetic energy while the energy in channel B is offset by an amount $\Delta$.  One can view $\Delta$ as the amount of energy needed to excite the two particles of type of A into those of type B.  Since for simplicity the dynamics are nonrelativistic in both channels the excitation energy does not effect the mass: it must be conserved and thus is the same in both channels. As will be discussed later, the restriction to non-relativistic dynamics is inessential.   The scattering problem consists of firing two particles of either type A or type B at each other and observing the outgoing particles which are either of the original type or through the interaction both converted to the other type.  This problem does not correspond to any interesting physical system.  However, it is a useful toy model for illustrating the relevant physics.

First consider the kinematic restrictions on $S$.  By Galilean invariance we expect the $S$-matrix that describes this physical process to conserve energy and momentum, and moreover to be independent of the total momentum.  We denote momentum of the two particles of species $j=A,B$ by $k_{j;1,2}$ where {1,2} labels particles 1 and 2 in the channel.  The total momentum $K$ is obviously
\begin{equation}
	K = k_{j;1}+k_{j;2}.
\end{equation}
Since we are considering a nonrelativistic problem, the energy can be written
\begin{equation}
	E = \oneover{2M}\left( k_{j;1}^{2} + k_{j;2}^{2}	\right) + \Delta_{j}
\end{equation}
where $\Delta_{j}=0$ for $j=A$ and $\Delta_{j}=\Delta$ for $j=B$.
It is easy to see that
\begin{equation}\label{eq:ekappa}
	E = \frac{K^{2}}{2(2M)} + \frac{\kappa_{j}^{2}}{2\mu} + \Delta_{j}
\end{equation}
where $\kappa_{j} = \half(k_{j;1}-k_{j;2})$ is the relative momentum of the two particles of species $j$, and $\mu = M/2$ is the reduced mass.  The asymptotic states that $S$ is kinematically allowed to connect, therefore, have relative momenta related by
\begin{equation}\label{eq:relative}
	\kappa_{k}^{2} = \kappa_{j}^{2} + 2 \mu \Delta_{jk},
\end{equation}
where we use the shorthand $\Delta_{jk} = \Delta_{j} - \Delta_{k}$.

Let us assume that the box is much larger than the range of the interaction between the particles.  Then, in large spatial regions the energy eigenstates in the box will approximate well the asymptotic form.  Because we are interested in periodic boundary conditions, outgoing asymptotic states will reach the box boundary and become incoming states on the opposite side of the box.  The energy eigenstates are asymptotically of the form
\begin{align}\label{eq:state}
	\ket{\Psi(E, K)}\asy = 	\sum_{j=A,B}	\Big(\phantom{+}& \alpha_{j} \ket{\text{in ;  $j$, } \kappa_{j},\ K} \\
									+	&			a_{j} 	\ket{\text{out;  $j$, } \kappa_{j},\ K}  \Big)	\nonumber
\end{align}
where the $\kappa_{j}$ are related to $E$ by \eqref{eq:ekappa} and ``in'' and ``out'' label the incoming and outgoing amplitudes respectively.  

Now let us consider the action of $S$ on asymptotic states.  It relates the coefficients by
\begin{equation}\label{eq:S}
	a_{j} = S(E,K)_{jk}\, \alpha_{k}.
\end{equation} 
Note the $S$ depends \emph{only} on the relative momenta; the dependence on $E$ and $K$ should be understood to be entirely through \eqref{eq:ekappa}.

Of course, only wavefunctions with certain energies will fit in a box of given size, and these energies will be determined by the dynamics encapsulated in $S$.  In position space, the pieces of this asymptotic wavefunction for channel $j$ are
\begin{widetext}
\begin{align}
	\braket{j, \, x_{1},\, x_{2}}{\text{in $j$, }\kappa_{j},\ K}\asy	&= \frac{e^{i K X}}{\sqrt{\kappa_{j}/\mu}}\left(	e^{i\kappa_{j} x}\, \theta(-x) +  e^{-i\kappa_{j} x}\, \theta(x)	\right)	\nn\\
	\braket{j, \,  x_{1},\, x_{2}}{\text{out $j$, }\kappa_{j},\ K}\asy 	&= \frac{e^{i K X}}{\sqrt{\kappa_{j}/\mu}}\left(	e^{i\kappa_{j} x}\, \theta(x) +e^{-i\kappa_{j} x}\, \theta(-x)	\right)	
\end{align}
\end{widetext}
where $X$ is the location of the center of mass $\half(x_{1}+x_{2})$ and $x=(x_{1}-x_{2})$ is the relative coordinate, $\theta$ is the usual Heaviside step function, and the normalization is such that the incoming probability flux and outgoing probability flux are the same if $S$ is unitary.  Here, ``asymptotic'' is understood to mean that the magnitude of the relative coordinate $x$ is much larger than range of the interaction.  The symmetric form reflects the fact that the particles are identical bosons.  

With these explicit asymptotic wavefunctions in hand, we note that  \eqref{eq:ekappa} does not fix the sign of $\kappa_j$. Either sign is acceptable since it can be reabsorbed in the definitions of $a_{j}$ and $\alpha_{j}$; by convention we define $\kappa_j$ to be positive.

Typical lattice simulations are performed with either periodic or twisted boundary conditions.  So far as the issues of principle addressed in this paper are concerned, one may verify that twisted conditions offer no additional leverage for the problem of extracting $S$ in this toy problem.  In any event, they may be included with minimal effort.  For simplicity we restrict our attention to simple periodic boundary conditions.  If the box under consideration has spatial extent $L$, periodic boundary conditions imply that the wavefunction obeys
\begin{equation}\label{eq:pbc}
	\braket{-\lbh,\, x_{2}}{\Psi(E,K)} = \braket{\lbh,\, x_{2}}{\Psi(E,K)}.
\end{equation}
In particular, we can project this equality onto each channel and require the equality to hold separately for the pieces which depend on $\exp(i k_{j;1}x_{2})$ and on $\exp(i k_{j;2}x_{2})$.  The $k_{j;1}$-dependent equality implies
\begin{equation}\label{cond1}
	a_{j} = \alpha_{j} e^{i k_{j;2}L}
\end{equation}
while the $k_{j;2}$-dependent equality implies
\begin{equation}\label{cond2}
	a_{j} = \alpha_{j} e^{-i k_{j;1}L}.
\end{equation}
By comparing these two equations we see that all channels imply the same quantization condition,
\begin{equation}\label{eq:Z}
	KL = 2\pi Z \, ;
\end{equation}
the integer $Z$  characterizes the total momentum.

Note that in lattice simulations one can restrict attention to correlation functions of operators with a fixed known momentum and thus extract energy levels with a fixed known $Z$.  We will assume that this can be done for this toy problem as well.  In principle one can try extracting the \Smatrix\ using a set of measurements at any fixed $Z$ or a mixed set of $Z$s.  Again, as far as issues of principle are concerned, this freedom gives no additional power for extracting information about $S$.  We will henceforth assume $Z$ is fixed and suppress dependence on the center of mass momentum $K$.

The relations between the coefficients of asymptotic incoming and outgoing wavefunction pieces for an energy eigenstate in Eqs.~(\ref{cond1}) and (\ref{cond2}), along with \eqref{eq:S}, imply
\begin{equation}\label{eq:geneig}
\left ( \BZL(E) -S(E) \right )\,\va = 0,
\end{equation}
where the boundary conditions were used to eliminate $a_{j}$ in favor of $\alpha_{j}$.  The matrix \BZL\ is diagonal in the channels and enforces the boundary conditions:
\begin{equation}
	\BZL(E)  =  (-1)^{Z}\left (  \begin{array}{cc}  \exp(- i \kappa_A L) & 0
	\\ 0 &  \exp(- i \kappa_B L)  \end{array}  \right ) .
\end{equation}
Note that \BZL\ depends on $E$ and $K$ only through the relative momenta and $Z$.

Equation~(\ref{eq:geneig}) is of the form of a generalized eigenvalue problem.  For boxes much larger than the size of the interaction, energy eigenstates exist for those energies for which the matrix $\left ( \BZL(E) -S(E) \right )$ has an eigenvector with eigenvalue zero.  Of course, if a matrix has a zero eigenvalue, it also has a determinant of zero.  Thus, the condition for an energy eigenstate for this toy problem to fit into a box of width $L$ is equivalent to
\begin{equation}\label{eq:Luscher_toy}
\det\Big[ \BZL(E) -S(E,Z) \Big] = 0.
\end{equation}

Equation (\ref{eq:Luscher_toy}) should be regarded as a generalized \Luscher\ formula for this multi-channel toy problem.  For any given \Smatrix\ one can use it to determine the energy levels.  Note that the result holds equally  well for the case where the dynamics are relativistic: the only difference is that the $\kappa_j$ are fixed by relativistic kinematics rather than non-relativistic kinematics.  

We note that even though \eqref{eq:Luscher_toy} was in a problem with only two channels, it immediately generalizes to any number of channels.  In particular, it also contains the familiar one-dimensional two-body single-channel \Luscher\ formula.

\section{A no-go theorem}\label{no-go-toy}

In this section we address the question of whether it is possible to fully reconstruct the \Smatrix\ for the toy problem from knowledge of the energy levels in finite boxes with periodic boundary conditions.  To make the problem concrete, we assume that we know the energy levels as a function of box size for a finite range of box sizes.  Given access to only such information about the dynamics, can one fully reconstruct the \Smatrix\  for this system at some given energy?  As will be shown, the answer is ``no''.

Suppose that all we knew about the dynamics for the two channel problem of the toy model is that it was quantum mechanical, then the only constraint we have for the \Smatrix\ is that it is unitary: $S \in U(2)$ at each energy.   The most general $U(2)$ matrix can be specified by 4 angle variables; a useful parameterization of $S$  is:
\begin{equation} \label{eq:param}
	S(E) = e^{i\phi(E)}e^{i \theta(E) \hat{n}(E)\cdot \vec{\sigma}}
\end{equation}
where $\vec{\sigma}$ represents the usual Pauli matrices and the energy-dependent unit vector $\hat{n}$ can be parameterized by
\begin{align}
	\hat{n} = &\sin(\alpha)\cos(\beta)\hat{x} +\sin(\alpha)\sin(\beta)\hat{y} + \cos(\alpha)\hat{z},
\end{align}
where we have suppressed the energy dependence through \eqref{eq:ekappa} of $\hat{n}$, $\alpha$, and $\beta$ for clarity.
Thus specifying $S(E)$ is equivalent to specifying $\phi$, $\theta$, $\alpha$ and $\beta$ as functions of energy; specifying the \Smatrix\ at one energy merely requires these functions' values at that energy.  Note, moreover, that the structure of Eq.~(\ref{eq:param}) is such that under the transformation 
\begin{align}\label{eq:trans}
\phi &\rightarrow \phi				&	\theta &\rightarrow 2 \pi - \theta \\
\alpha & \rightarrow \pi - \alpha 	&	\beta  &\rightarrow \beta + \pi \nonumber
\end{align}
the \Smatrix\ is unchanged.  This allows us to choose as a convention a  restricted range of the parameters:  $0 \le \phi < 2\pi$, $0 \le\theta < 2 \pi$, $0 \le \alpha < \pi$ and $-\pi/2 \le \beta < \pi/2$.

It is easy to show that the energy levels given by the generalized \Luscher\ formula of \eqref{eq:Luscher_toy} are completely independent of the parameter $\beta$.  This means that it is impossible to determine $\beta(E)$ from knowledge of the energy levels and hence impossible to determine the \Smatrix.  

To see why the energy levels do not depend on $\beta$,  consider the energy-dependent diagonal matrix 
\begin{equation}\label{eq:A}
D(E)=\exp(i \chi(E) \sigma_3)
\end{equation}
 where $\chi(E)$ is a general energy-dependent function that specifies $D(E)$.  From the multiplicative property of the determinant and the unitary nature of $D$, it follows that if $\det\left [ \BZL(E) -S(E) \right ]$ vanishes, then $\det\left [D^\dagger(E) \left(\BZL(E) -S(E)\right ) D(E) \right ]$ does too.  Moreover, since $\BZL(E)$ is diagonal it commutes with $D(E)$.  Thus, whenever \eqref{eq:Luscher_toy} holds for \Smatrix\ $S(E)$ for some given box size $L$, it also holds for 
 \begin{equation}
 S'(E) \equiv D^\dagger S(E) D.
\end{equation} 
It is easy to see that  $S'(E)$ is related to $S(E)$ in the following way:
\begin{align}\label{eq:prime}
\phi ' (E)	&= \phi(E) 		&	\theta '(E) & 	=	\theta(E) \\
\alpha '(E) &= \alpha(E)	&	\beta '(E)  &	= 	\beta(E) + 2 \chi(E) \, . \nonumber
\end{align}
Equation (\ref{eq:prime}) implies that the $\beta$ parameter can be shifted by an arbitrary energy-dependent amount without affecting the energy levels for any box size, provided that the boxes are large enough for the generalized \Luscher\ formula to hold.  Clearly the  the energy levels alone are insufficient to determine $\beta(E)$ and hence the \Smatrix\ cannot be fully specified.

It is worth noting that the parameter $\beta$ is associated with time-reversal-noninvariant dynamics: for $\beta=0$ the \Smatrix\ is symmetric but otherwise the \Smatrix\ is generically not symmetric.  It is well known\cite{PhysRev.89.619} that the \Smatrix\ is symmetric for theories in which the dynamics is \T-symmetric.

Thus, while there is a no-go theorem for fully determining the \Smatrix\ for the most general case of dynamics, it may appear that it is evaded if one knows that the dynamics of the system is \T-symmetric since  for \T-symmetric theories one can simply set $\beta$ to zero at the outset.  However, that is not the case.  There remains a discrete ambiguity even in the \T-symmetric case. Suppose that we begin with an \Smatrix\ $S(E)$ with $\beta=0$ that satisfies the generalized \Luscher\  formula.  Next, we pick $D$ by choosing $\chi(E)$ to be $-\pi/2$ for all $E$.  Such a transformation keeps \Smatrix\ symmetric.  In that case we know from the argument above that $S'(E) \equiv D^\dagger S(E) D$ will have identical energies for all boxes large enough for the generalized \Luscher\ formula to apply.   Using \eqref{eq:prime} and \eqref{eq:trans} one sees that relationship between the parameters of $S'(E)$ and $S(E)$ is given by
\begin{align}
\phi ' (E)&= \phi(E) 		&	\theta '(E) &=	2\pi-\theta(E) \\
\alpha '(E) &=\pi-\alpha(E) &	\beta '(E)  &= 0.\nonumber
\end{align}
As advertised this is a valid symmetric matrix with $\beta=0$ which is physically distinct from $S(E)$ provided that both $\theta$ and $\alpha$ are nonzero.  The nature of this discrete ambiguity is that of a sign ambiguity for the off-diagonal matrix elements of $S$.

\section{physical relevance of sign indeterminacies}\label{no-go-toy}

One might question whether the sign ambiguity described above can be eliminated by a choice of convention and, more importantly, whether or not it corresponds to anything physical.  Indeed, it has been argued that the inability to determine the sign in question does not put restrictions on the determination of physical observables since it is only tied to an arbitrary phase in the wave function of the two channels with respect to each other\cite{Oset:2012}.   However, this critique is correct only to the extent that there are no observables of interest which depend on the interference between the two outgoing channels.  From a quantum information perspective there is real information contained in the sign as there is no reason in principle that observables sensitive to interference between the channels cannot be measured.

The sensitivity of the $S$-matrix to the sign is {\it not} merely one of philosophical interest.  There are directly measurable, physical consequences associated with the sign of the off-diagonal matrix element (or more generally, its analog in more complicated examples).  These can be seen most easily in cases where there is a symmetry in the problem which causes the $S$-matrix to break into symmetry sectors.   Each sector can be calculated separately on the finite box and the $S$-matrix for each sector will have the type of sign ambiguity discussed here provided it has two or more channels (an example in 3+1 dimensions is rotational symmetry which leads to a partial wave analysis.)   Note that the incident state in physical scattering experiments does not correspond to any single partial wave; rather it is well-approximated by a plane wave and plane waves mix all partial waves in a manner in which their relative phases are fixed.  Moreover, differential cross sections, the basic observable in scattering experiments, superpose amplitudes in multiple partial waves and through this superposition are sensitive to the analogous \emph{relative} sign ambiguity which arises in each partial wave.  

\subsection{An example in 1+1 dimensions}
  
This point is illustrated most simply by again turning to 1+1 dimensions.  In this case, if the symmetry is parity (as opposed to rotational symmetry in 3+1 dimensions), the analog of the various partial waves are distinct parities of the states.  That is, there are two possible ``partial waves'': even and odd.  However, the scattering experiments of interest  involves particles being shot in from  the left or right and thus correspond to superpositions of ``partial waves''.  

In the example studied so far we considered identical bosons; this restricted consideration to  the even parity sector.   Thus was done largely for convenience so that we did not have to consider  the role of the odd-parity sector.  Dropping this restriction, let us again imagine a nonrelativistic two-channel scattering process involving pairs of  particles.  In channel $A$ there are two particles (denoted $A_1$ and $A_2$); in channel $B$ there are also two particles ($B_1$ and $B_2$).   The dynamics are parity symmetric and can induce transitions between the two channels ({\it eg.}  $A_1 A_2 \rightarrow B_1 B_2$).    As in our previous example, there is an energy offset  of $\Delta$ between the two channels.  Also, as in the previous example, we will assume that the two particles in each channel, when isolated, are degenerate in energy.  However, unlike the previous example, here  we will take them to be distinguishable.  

The \Smatrix\ is  four-by-four ---there are two channels in the sense of $A$-$B$ and two spatial sectors which we can take to be even and odd.  Due to parity, the dynamics of the scattering processes separates into  a reducible $S$-matrix representation associated with the even and odd Hilbert subspaces (i.e. subspaces spanned by even- and odd-parity  wavefunctions). Thus the  $S$-matrix in $2\times2$ block form is 
\begin{equation}
S^{\rm P}(E) = \begin{pmatrix} S^{+}(E) &\vline & \mathbb{0}\\
\hline 
\mathbb{0} & \vline & S^{-}(E) \end{pmatrix}
\end{equation}
where the $+/-$ superscripts indicate, respectively, $S$-matrices for even and odd sectors and the superscript P indicates that this is in a basis of good parity.  $S^{\pm}(E)$ are  two-by-two matrices describing the $A$-$B$ basis; they are symmetric due to time reversal invariance.

By studying the energy levels in a finite box for even states and odd states separately one can attempt to extract $S^{\pm}(E)$.  Indeed for $S^{+}$ the algorithm for doing so is identical to the case of bosons considered earlier.  For  $S^{-}$  the only change is that the matrix $B$ acquires an overall minus sign.  In both cases one is unable to determine the sign of the off-diagonal matrix element.  

To show that these signs have physical content, it is useful to transform the \Smatrix\ from a parity basis to a right-left basis which is the one associated with physical scattering.  In the right-left basis the upper (lower) blocks describe the cases where particles of type one are incident moving to the the right (left) (i.e. \emph{from} the left (right).)  The blocks on the left (right) correspond to outgoing particle of type one moving to the right (left).  The transformation to the right-left basis is given by the unitary matrix 
\begin{equation}
A = \sqrt{\frac{1}{2}} \begin{pmatrix} \mathbb 1 &\vline &\mathbb 1 \\
\hline 
\mathbb 1& \vline & -\mathbb 1
\end{pmatrix} \; .
\end{equation}
Thus, the \Smatrix \, in the right-left basis is
\begin{equation} \label{RL}
\begin{split}
S^{\rm RL} &=A^\dagger S^{\rm P}(E) A \\&= \frac{1}{2}\begin{pmatrix} S^{+}(E) + S^{-}(E) & \vline & S^{+}(E) -S^{-}(E) \\
\hline 
S^{+}(E) - S^{-}(E) & \vline & S^{+}(E) + S^{-}(E) \end{pmatrix}
\end{split}
\end{equation}

Let us consider a particular physical process.  Particles of type $A$ are shot in with particle one moving to the right and particle two moving to the left with some fixed energy and upon scattering they can turn into particles of type $B$.  Particle $B_1$ can come out moving to either to the right or the left. Let us focus on the physical question of whether a created particle of type $B$ is more likely to come out moving to the left or the right. From Eq.~(\ref{RL}), the probability that this experiment will yield a particle of type $B_1$ moving to the right minus the probability that it yields $B_1$ moving to the left is given by
\begin{equation}
P_{R}-P_{L}= {\rm Re} \left( S^{+}_{AB}(E)\, S^{-}_{AB}(E)^*  \right) 
\end{equation}
Note that this probability difference will flip if either $S^{+}_{AB}(E)$ or $ S^{-}_{AB}(E)$ flips sign (except in the exceedingly unlikely case where one of them is {\it exactly} zero or their complex phases differ by {\it  exactly} $\pi/2$ when the probability of the directions is accidentally equal).  Thus the prediction of whether particle $B_1$ is more likely to come out moving to the left or the right will flip if either $S^{+}_{AB}(E)$  and $ S^{-}_{AB}(E)$ or flips signs.    This conclusively demonstrates that there is extractable physical information describing scattering that is contained in the signs of  $S^{+}_{AB}(E)$  and $ S^{-}_{AB}(E)$.   However, neither of these signs is extractable from studies of energies levels in a finite box.

\subsection{Examples in 3+1 dimensions}

Of course, the 1+1 dimensional example is merely illustrative.  The same sensitivity occurs in 3+1 dimensions.  Consider as a simple example the reaction $\pi^+ \pi^- \rightarrow K^+ K^-$.  Right at threshold this is dominated by the $s$-wave, $I=0$ sector.  This reduces the problem to two channels---$\pi-\pi$ and $K-K$.  An analog of Eq.~(\ref{eq:Luscher_toy}) has been derived; again there is an intractable sign ambiguity in determining the off-diagonal matrix element for the \Smatrix\  from spectral information; consequently the sign of the transition amplitude is not fixed \cite{Hansen:2012tf,Oset:2012}.  The sign of this amplitude is irrelevant if one is only interested in the cross-section of this process at threshold, since only the squared modulus of the amplitude is relevant.  

As in 1+1 dimensions, the sign does carry important physical information.  To see this, consider what happens in the $\pi^+ \pi^- \rightarrow K^+ K^-$ reaction slightly above threshold.  In that case, one might expect a small, but non-negligible, contribution from the $p$-wave, $I=1$ sector.  This sector also has two channels---$\pi-\pi$ and $K-K$---and again one could derive an an analog of Eq.~(\ref{eq:Luscher_toy}).  Again, the sign of the off-diagonal off-diagonal matrix element for the \Smatrix\ in this sector cannot be determined from  from spectral information in a finite box.

In this case, the differential cross-section to leading order in the $p$-wave amplitude is given by:
\begin{equation}
\left . \frac{d \sigma}{d \Omega} \right |_ {\pi^+\pi^-\rightarrow K^+K^- } = |A_s|^2 + 2 {\rm Re} \left (A_s^*A_p\right) \cos(\theta)
\end{equation}
where the angle $\theta$ is the angle of the outgoing $K^+$ relative to the incident $\pi^+$, the $s$- and $p$-wave amplitudes and the coefficients $A_s$ and $A_p$ depend on kinematical factors and the appropriate Clebsch-Gordan coefficients to combine the channels, and are proportional to the off-diagonal matrix elements of the two sectors.  Since they are proportional to the off-diagonal matrix elements, their sign cannot be determined from the spectrum in finite boxes.  However, if either $A_s$ or $A_p$ changes sign, the differential cross-section changes---indeed whether the scattering preferentially has the $K^+$ produced in a forward or backward direction flips if {\it either} sign is flipped.


\subsection{General considerations}

These examples in 1+1 and 3+1 dimensions demonstrate that the information contained in the sign of the off-diagonal S-matrix elements comes through interference with some other sector.  The key point is the relative signs in the two sectors carry information.  

In determining the S-matrix from an underlying theory one must one must typically make choices of convention.   In a field-theoretic setting, two-to-two $S$-matrices  can be obtained from an LSZ reduction from the four-point correlation function \cite{SC} of operators carrying the quantum numbers of the particles of interest along with the two-point correlation functions for each type of operator.   Consider the process $A+B \rightarrow C+D$.  As a first step one needs to identify operators with the correct quantum numbers (of the particles of interest.)  For simplicity, these operators can be normalized in a manner that their two-point correlation functions satisfy:
\begin{align}
\lim_{q^2\goesto M_A^2} & G_A^{-1}(q;M_A) \Pi_A(q)  \rightarrow 1 \\
\Pi_A(q) & \equiv \int \frac{d^4 x}{(2 \pi)^4 } e^{i q_\mu x^\mu} \langle T ( \overline{A}(x) A(0) ) \rangle
 \end{align}
where $G_A^{-1}(q;M_A)$ is the propagator for a non-interacting particle with quantum number of $A$ with the mass $M_A$.   The scattering amplitude is then given in terms of the 4-point connected correlation function, $ \Gamma_{ABCD}(k^A,k^B,k^C,k^D)$: 
\begin{widetext}
\begin{align}
&\mathcal{A} =G_A^{-1}(k^A;M_A) G_B^{-1}(k^B;M_B) G_C^{-1}(k^C;M_C)G_D^{-1}(k^A;M_D) \Gamma_{ABCD}(k^A,k^B,k^C,k^D) \; \; \;  {\rm where} \label {LSZ}\\
&\Gamma_{ABCD}(k^A,k^B,k^C,k^D)=\int \frac{d^4 x_A}{(2 \pi)^4 } \frac{d^4 x_B}{(2 \pi)^4 } \frac{d^4 x_C}{(2 \pi)^4 }  \frac{d^4 x_C}{(2 \pi)^4 }  e^{i k^A_\mu x_A^\mu} e^{i k^B_\mu x_B^\mu}
e^{-i k^C_\mu x_C^\mu} e^{-i k^D_\mu x_D^\mu} \left \langle T \left ( \overline{D}(x_D) \overline{C}(x_C) B(x_B) A(x_A) \right )\right  \rangle_c \nonumber
\end{align}
\end{widetext}
 This procedure involves a  choice of convention, namely the choice of operator with the quantum numbers of interest.  For example, one can choose the operator with pion quantum numbers to be $-i \overline{q} \gamma_5 \vec{\tau} q$ rather than  $i \overline{q} \gamma_5 \vec{\tau} q$ and with the standard LSZ procedure, it will yield the opposite sign for $T$-matrix elements for processes involving a single pion.

The fact that the sign may depend on convention does {\it not} mean that it contains no information; as demonstrated above the sign is physically relevant.  The key point   is  that once the conventions are specified, there is real content in the sign of the resulting $S$-matrix.  The critical thing  is that the choice of convention once made needs to be applied {\it consistently} across the entire theory.  Thus, for example one may select any choice of conventions desired for the different partial wave sectors of a process, but the only way to consistently recover the correct differential cross section is to choose the same convention for all sectors.   Of course, in computing the energy levels in a box, one can make no choices of convention for the $S$-matrix and thus it is not surprising that one cannot fix signs which depend on such conventional choices. 

In the examples discussed so far, the physical information resided in the {\it relative} signs in  various partial waves.  Clearly, in terms of direct scattering observables, this is where the sign information manifests itself.  One might ask the question of whether there is any information in the absolute sign for all the partial waves.  In this context, it is noteworthy  that  there are cases  in which  the standard LSZ procedure has no ambiguities and and the absolute sign of the off-diagonal $S$-matrix element is fixed.     This occurs  when  an  incident particle-antiparticle  pair interacts and turns into a different particle, anti-particle pair, for example $\pi^+ \pi^- \rightarrow K^+ K^-$.  From, Eq.~(\ref{LSZ}), it is clear that any choice of operators   will yield the same sign for this process.  

There is an alternative way to see that the absolute sign of the amplitude $\pi^+ \pi^- \rightarrow K^+ K^-$ is fixed, namely from crossing symmetry and analyticity of the scattering amplitude.  From general field-theoretic considerations, the $S$-matrix can be considered as a  general function of the kinematic variables even when they are analytically continued off-shell---indeed the scattering amplitude in Eq.~(\ref{LSZ}) is constructed to apply off-shell.   This function is known to have crossing symmetry so that the function which describes $\pi^+ \pi^- \rightarrow K^+  K^-$ also describes   $\pi^+  K^-  \rightarrow \pi^+  K^-$; it is merely evaluated a different kinematic point \cite{SC}.  Moreover, the two kinetic regions can be shown to both be in a common region which is analytic \cite{SC}.  This in turn implies that one can not flip the sign of the scattering amplitude of  $\pi^+ \pi^- \rightarrow K^+  K^-$ without also flipping the sign of the amplitude of $\pi^+  K^-  \rightarrow \pi^+  K^-$.  

 On the other hand, the process $\pi^+  K^-  \rightarrow \pi^+  K^-$ is elastic.   As such the form of the scattering amplitude at threshold is fixed:
\begin{equation}
{\cal A} =\frac{2 \pi}{\mu}\, \frac{1}{-1/a - i k}= \frac{2 \pi}{\mu}  \left(	\frac{-a + i a^2 k} {1 + a^2  k^2}	\right)
\end{equation}
where $\mu$ is the reduced mass, $k \equiv \sqrt {2 \mu E}$ where $E$ is the kinetic energy in the center of mass and $a$ is the scattering length.  Note that there is no overall sign ambiguity in the scattering amplitude--the imaginary part of the amplitude is fixed to be positive.  Since there is no freedom to flip the sign of the amplitude  $\pi^+  K^-  \rightarrow \pi^+  K^-$  there is also no freedom to flip the sign of  the amplitude for $\pi^+ \pi^- \rightarrow K^+  K^-$.  This is significant in that the {\it absolute} sign of the off-diagonal terms in the $S$-matrix are fixed---not merely the relative signs of partial waves.   While the absolute sign does not affect the scattering observables, it does encode important physics.  This physics cannot be obtained from the energy levels in a box.  Moreover, even in cases where crossing symmetry does not relate the process to an elastic one and fix the absolute sign for the off-diagonal terms in the $S$-matrix,  that absolute sign for a process still contains important quantum information as it is fixed relative  to the absolute sign of the crossed process.

\subsection{Practical Implications}

As noted above, both the absolute sign and the  relative sign of the various partial waves of the off-diagonal $S$-matrix contain important physical information.  If one's sole concern is computing scattering observables then only the relative signs matter.   This is important for scattering in 3+1 dimensions in that {\it some} information about the relative signs may be extractable from energy levels in the box.  To see how, consider the form of Eq.~(\ref{eq:Luscher_toy}).  This general form holds for three spatial dimensions as well as one.    The key point concerns symmetry: to the extent that the $S$-matrix and $B(L)$, the matrix encoding the boundary conditions share symmetries, one can only obtain information about the relative signs of different partial waves within any symmetry class from energy levels in a box.  The relative signs between different symmetry classes cannot be obtained since the energy levels of each symmetry class are unaffected by other symmetry classes.  Two things are noteworthy.  The first is that since the symmetry of the box differs from the spherical symmetry of the $S$-matrix, the symmetry classes are for the smaller class of symmetries---{\it i.e.} that of the box.  This in turn means that Eq.~(\ref{eq:Luscher_toy}) automatically connects certain partial waves (for example $L=0$ and $L=4$) and, to the extent that one can obtain enough accurate information doing measurements of energy levels in the box to learn about both the $L=0$ and $L=4$ partial waves, one can about the relative sign of the two.   The second is that by studying energy levels for $\vec{K} \ne 0$, one reduces the symmetry of $B$ and induces additional mixing between partial waves, including for example between $s$-wave and $p$-wave \cite{rummukainen1995, doring2012}.  This means that if one can do sufficiently accurate information for studies using $\vec{K} \ne 0$ one can fix the relative sign of the off-diagonal matrix elements for those partial waves which are determinable from $\vec{K} \ne 0$ measurements.

Thus, the principal practical implication of the no-go theorem for the direct computation of scattering observables is that one {\it must} do calculations at $\vec{K} \ne 0$ in order to fix the relative signs.  This can greatly add to the complexity of the calculation as it means that one cannot fully exploit the symmetries of the problem.   Moreover, the information one can obtain from lattice studies is necessarily limited.  Thus, for example, one might have have the computational resources to compute the $s$-wave and $p$-wave scattering and their relative sign, one still does not know their sign relative to the $d$-wave scattering even if one has the computational resources to extract the $d$-waves on their own.

Finally, it should be noted that one of the reasons that one wants to {\it ab initio} calculations from QCD is to understand processes at a deep level.  The quantum mechanical $S$-matrix contains information beyond its immediate connection to the scattering observables.  As an off-shell function it is connected to other processes, including the crossed process.  Ideally one wants access to as much of this information as possible.  This no-go theorem means that some of this information, even on-shell is not obtainable from energy levels in a box since the relative  sign of a process and its crossed process are fixed.

\begin{acknowledgements}
	The authors thank U.S. Department of Energy for support under grant \#DE-DG02-93ER-40762. E.~B. also thanks Jefferson Science Associates for support under the JSA/JLab Graduate Fellowship program. We thank Thomas Loomis for valuable discussions.  We thank Eulogio Oset and Zohreh Davoudi for useful discussions and in particular for pointing out that an earlier version of this work did not demonstrate the physical significance of the sign of the off-diagonal matrix element.  We also thank Eulogio Oset for pointing out the role of $\vec{K} \ne 0$ in mixing partial waves.
\end{acknowledgements}


\end{document}